\UseRawInputEncoding

\documentclass[11pt]{article}
\usepackage{longtable,supertabular}
\usepackage{amsmath}
\usepackage{amssymb}
\usepackage{latexsym}
\usepackage{cite}

\title{\bf Covering $b$-Symbol Metric Codes and the Generalized Singleton Bound}
\author{Hao Chen
  \thanks{Hao Chen is with the College of Information Science and Technology/Cyber Security, Jinan University, Guangzhou, Guangdong Province, 510632, China, haochen@jnu.edu.cn. The research of Hao Chen was supported by NSFC Grant 62032009.}}

\begin{document}

\maketitle
\begin{abstract}
Symbol-pair codes were proposed for the application in high density storage systems, where it is not possible to read individual symbols. Yaakobi, Bruck and Siegel proved that the minimum pair-distance $d_2$ of binary linear cyclic codes satisfies $d_2 \geq \lceil 3d_H/2 \rceil$ and introduced $b$-symbol metric codes in 2016. In this paper covering codes in $b$-symbol metrics are considered. Some examples are given to show that the Delsarte bound and the Norse bound for covering codes in the Hamming metric do not hold true for covering codes in the pair metric. We give the redundancy bound on covering radius of linear codes in the $b$-symbol metric and give some optimal codes attaining this bound. Then we prove that there is no perfect linear symbol-pair code with the minimum pair-distance $7$ and there is no perfect $b$-symbol metric code if $b\geq \frac{n+1}{2}$. Moreover a lot of cyclic and algebraic-geometric codes are proved non-perfect in the $b$-symbol metric. The covering radius of the Reed-Solomon code in the $b$-symbol metric is determined. As an application the generalized Singleton bound on the sizes of list-decodable $b$-symbol metric codes is also presented. Then an upper bound on lengths of general MDS symbol-pair codes is proved.\\

\end{abstract}

\section{Introduction}

The Hamming weight $wt({\bf a})$ of a vector ${\bf a} \in {\bf F}_q^n$ is the number of non-zero coordinate positions. The Hamming distance $d_H({\bf a}, {\bf b})$ between two vectors ${\bf a}$ and ${\bf b}$ is the Hamming weight of ${\bf a}-{\bf b}$. The minimum Hamming distance of a code ${\bf C} \subset {\bf F}_q^n$,  $$d_H({\bf C})=\min_{{\bf a} \neq {\bf b}} \{d_H({\bf a}, {\bf b}): {\bf a} \in {\bf C}, {\bf b} \in {\bf C} \},$$  is the minimum of Hamming distances $d_H({\bf a}, {\bf b})$ between any two different codewords ${\bf a}$ and ${\bf b}$ in ${\bf C}$. Theory of Hamming error-correcting codes has been fully developed and numerous constructions have been proposed, we refer to \cite{MScode,HP,TV,Lint1}. The Singleton bound for a linear $[n, k, d]_q$ code is $d_H \leq n-k+1$. When the equality holds, this code is an MDS code.  Let ${\bf F}_q$ be an arbitrary finite field, $P_1,\ldots,P_n$ be $n \leq q$ distinct elements in ${\bf F}_q$. The Reed-Solomon code is defined by $$RS(n,k)=\{(f(P_1),\ldots,f(P_n)): f \in {\bf F}_q[x],\deg(f) \leq k-1\}.$$ This is an $[n, k, n-k+1]_q$ linear MDS code attaining the Singleton bound, since a degree $\deg(f) \leq k-1$ nonzero polynomial has at most $k-1$ roots. \\

A general code ${\bf C} \subset {\bf F}_q^n$ is called $(d, L)$ list-decodable in the Hamming metric if each ball $B_H({\bf x}, d)=\{{\bf y}: d_H({\bf x}, {\bf y}) \leq d\}  \subset {\bf F}_q^n$,  contains at most $L$ codewords of ${\bf C}$ for each ${\bf x} \in {\bf F}_q^n$, see \cite{ShangguanTamo}.  The generalized Singleton bound $$|{\bf C}| \leq L q^{n-\left \lfloor\frac{(L+1)d}{L}\right \rfloor} $$ for $(d, L)$ list-decodable Hamming metric codes was proved in \cite{ShangguanTamo}. When $d=\left \lfloor \frac{d_H({\bf C})-1}{2} \right \rfloor$ and the list size $L=1$, this is the classical Singleton bound.\\

For a general code ${\bf C} \subset {\bf F}_q^n$, we define its covering radius in the Hamming metric by $$R_H({\bf C})=\max_{{\bf x} \in {\bf F}_q^n} \min_{{\bf c} \in {\bf C}} \{wt({\bf x}-{\bf c})\}.$$ Hence the Hamming balls $B(x, R_H({\bf C}))$ centered at all codewords $x \in {\bf C}$, with the radius $R_H({\bf  C})$ cover the whole space ${\bf F}_q^n$, and moreover this radius is the smallest possible such radius. For this topic of coding theory, we refer to the book \cite{CHLL}. \\

A code in the Hamming metric is perfect if the covering radius of this code is $R_H({\bf C})=\lfloor (d_H({\bf C})-1)/2 \rfloor$. For a length $n$ perfect code ${\bf C}$, the whole space ${\bf F}_q^n$ is the disjoint union of the Hamming balls of the radius $R_H({\bf C})=\lfloor (d_H({\bf C})-1)/2 \rfloor$ centered at all codewords. This is an intersecting point of both packing and covering problems in the Hamming metric space ${\bf F}_q^n$. Perfect codes in the Hamming metric have the same parameters as parameters of Hamming codes, or the binary $[23, 12, 7]_2$ Golay code or the ternary $[11, 6, 5]_3$ Golay code, see \cite{Lint,Tie1973,CHLL}. We cite the comment in page 49 of \cite{HP}, " The classification of perfect codes as summarized in this theorem was a significant and difficult piece of mathematics." Covering codes and in particular perfect codes in the rank-metric have been studied in \cite{BR17,KChen,GY3}. Another interesting intuition of the generalized covering radii of linear codes was introduced and studied in a recent paper \cite{EFS}.\\

Hence it is interesting to study perfect codes in $b$-symbol metrics. There have been extensive research on covering radii of codes in the Hamming metric. For example, in the paper \cite{DJ1991}, covering radii of more than six thousand binary cyclic codes were calculated and determined. On the other hand many MDS symbol-pair codes have been constructed by cyclic codes or constacyclic codes, see \cite{KZL15,Dinh,Dinh20,CLL2017,ML22}. Then what will happen when these well-known linear codes such as cyclic codes, constacyclic codes, Reed-Solomon codes and algebraic-geometric codes are considered as covering $b$-symbol metric codes?\\

Symbol-pair codes were introduced for high density data storage, we refer to \cite{CB11,CL2011}. Set $({\bf F}_q^2)^n=\{( x_1, x_2), ( x_2, x_3), \ldots, (x_{n-1},  x_n), (x_n, x_1)): {\bf x}=( x_1, \ldots, x_n) \in {\bf F}_q^n\}.$ The space $({\bf F}_q^b)^n$ for $b=3, 4, \ldots, n-1$ can be defined similarly. The pair metric on ${\bf F}_q^n$ is defined as follows. For ${\bf x}=(x_1, \ldots, x_n) \in {\bf F}_q^n$, the mapping $\pi_2: {\bf F}_q^n \longrightarrow ({\bf F}_q^2)^n$ is defined by  $\pi_2 ({\bf x})=((x_1,x_2), \ldots, (x_n, x_1)) \in ({\bf F}_q^2)^n$. The pair weight $wt_2({\bf x})$ of ${\bf x} \in {\bf F}_q^n$ is $$wt_2({\bf x})=wt(\pi_2({\bf x}))=|\{i: (x_i, x_{i+1}) \neq 0 \}|.$$ The pair-distance $d_2$ is defined by $$d_2({\bf x}, {\bf y})=wt_2(\pi_2({\bf x})-\pi_2({\bf y})),$$ see \cite{CB11,CL2011}. It follows from the definition that $$\min \{d_H({\bf x}, {\bf y})+1, n\} \leq d_2({\bf x}, {\bf y}) \leq \min\{2d_H({\bf x}, {\bf y}),n\}.$$ The ball in the pair-metric is $$B_2({\bf x}, r)=\{{\bf y} \in {\bf F}_q^n: d_2({\bf x}, {\bf y}) \leq r\}.$$ For a code ${\bf C} \subset {\bf F}_q^n$, the minimum pair-distance is $$d_2({\bf C})=\min \{d_2({\bf x}, {\bf y}): {\bf x} \neq {\bf y}, {\bf x}\in {\bf C}, {\bf y} \in {\bf C}\}.$$ Then we have $$\min\{d_H({\bf C}+1, n\} \leq d_2({\bf C}) \leq \min\{2d_H({\bf C}), n\}.$$\\

The Singleton bound $|{\bf C}| \leq q^{n-d_2+2}$ for symbol-pair codes was proved in \cite{CJKWY13} and some MDS symbol-pair codes have been constructed in \cite{CJKWY13,KZL15,CLL2017,DG,Dinh,ML22}. In \cite{YBS16} $b$-symbol metric was introduced and the following lower bound of the minimum pair distances for linear binary cyclic codes was proved, $$d_2 \geq \left \lceil \frac{3d_H}{2} \right \rceil.$$ A general symbol-pair code ${\bf C} \subset {\bf F}_q^n$ is called $(d_{list}, L)$ list-decodable if each ball $B_2({\bf x}, d_{list}) \subset {\bf F}_q^n$,  contains at most $L$ codewords of ${\bf C}$ for each ${\bf x} \in {\bf F}_q^n$. The list-decodability and list-decoding of symbol-pair codes have  been studied in \cite{LXY1}.  More constructions and bounds for symbol-pair codes was given in \cite{EGY20}. Generalized pair weights of linear codes were introduced and studied in \cite{LP}. \\

We recall the $b$-symbol metric on ${\bf F}_q^n$ for $2 \leq b \leq n-1$.  For ${\bf x} \in {\bf F}_q^n$, set $\pi_b({\bf x})=((x_1, \ldots, x_b), (x_2, \ldots, x_{b+1}), \ldots, (x_n, x_1, \ldots, x_{b-1}) )\in ({\bf F}_q^b)^n$. Then the $b$-symbol metric on ${\bf F}_q^n$ was introduced in \cite{YBS16}. The $b$-symbol weight $wt_b({\bf x})$ of ${\bf x}=(x_1, \ldots, x_n) \in {\bf F}_q^n$ is $$wt_b({\bf x})=wt(\pi_b({\bf x}))=|\{i: (x_i, x_{i+1}, \ldots, x_{i+b-1})\neq 0\}|.$$ The $b$-symbol distance is defined by $$d_b({\bf x}, {\bf y})=wt_b(\pi_b({\bf x})-\pi_b({\bf y})).$$ The $b$-symbol ball in ${\bf F}_q^n$ is $$B_b({\bf x}, r)=\{{\bf y} \in {\bf F}_q^n: d_b({\bf x}, {\bf y}) \leq r\}.$$ It is clear that $$\min\{d_H({\bf x}, {\bf y})+b-1, n\} \leq d_b({\bf x}, {\bf y}) \leq \min\{bd_H({\bf x}, {\bf y}),n\}, $$ see Proposition 9 in \cite{YBS16}. The minimum $b$-symbol distance of a code ${\bf C}$ is $$d_b({\bf C})=\min\{d_b({\bf x}, {\bf y}): {\bf x} \neq {\bf y}, {\bf x}\in {\bf C}, {\bf y} \in {\bf C}\}.$$ For recent work on minimum $b$-symbol distances of linear cyclic codes, we refer to \cite{Dinh20,SOS21}.\\

For a general code ${\bf C} \subset {\bf F}_q^n$, we define its covering radius in the $b$-symbol metric by $$R_b({\bf C})=\max_{{\bf x} \in {\bf F}_q^n} \min_{{\bf c} \in {\bf C}} \{wt_b({\bf x}-{\bf c})\}.$$ Hence the balls $B_b({\bf x}, R_b({\bf C}))$ centered at all codewords $x \in {\bf C}$, with the radius $R_b({\bf  C})$ cover the whole space ${\bf F}_q^n$, and moreover this radius is the smallest possible such radius. To the best of our knowledge there has been no previous work on covering codes in the $b$-symbol metric. It is interesting to determine perfect codes in the $b$-symbol metric, that is these codes ${\bf C} \subset {\bf F}_q^n$ satisfying $$R_b({\bf C})=\left \lfloor \frac{d_b({\bf C})-1}{2} \right \rfloor.$$ A code ${\bf C} \subset {\bf F}_q^n$ is called $(d_{list}, L)$ list-decodable $b$-symbol metric code if the ball $B_b({\bf x}, d_{list}) \subset {\bf F}_q^n$,  contains at most $L$ codewords of ${\bf C}$ for each ${\bf x} \in {\bf F}_q^n$. There is no previous work about list-decodable codes in the $b$-symbol metric.\\

There are many  results about covering codes in the Hamming metric, for example, the Delsarte bound in \cite{Delsarte} and the Norse bound in \cite{HKM78}. In Scetion 3, we give some examples, which show that these bounds do not hold for covering codes in the pair metric. Then the redundancy upper bound on covering radii of linear codes in the $b$-symbol metric is given and some optimal codes attaining this bound are constructed. The covering radius of the Reed-Solomon code in the $b$-symbol metric is determined. We prove that the perfect $b$-symbol metric code does not exist when $b \geq \frac{n+1}{2}$ and the perfect linear symbol-pair code of the minimum pair-distance $7$ does not exist neither.  Moreover it is proved that a linear $[n, k]_q$ code can not be perfect in the $2(k+1)$-symbol metric.  Many well-known cyclic codes, constacyclic codes and algebraic-geometric codes are shown to be not perfect in the pair metric. On the other hand the generalized Singleton bound on list-decodable $b$-symbol metric codes is proved as an application of covering $b$-symbol metric codes. From the generalized Singleton bound on symbol-pair codes, we give an upper bound on the lengths of a MDS symbol-pair codes.

\section{The covering codes in the $b$-symbol metric}

We prove the following two upper bounds on the covering radius of a linear code in ${\bf F}_q^n$ in the $b$-symbol metric.\\

{\bf Proposition 2.1.} {\em Let ${\bf C} \subset {\bf F}_q^n$ be a general code. Then $\min \{R_H({\bf C})+b-1, n\} \leq R_b({\bf C}) \leq \min\{b R_H({\bf C}), n\}$.}\\

{\bf Proof.} From the inequality $\min\{d_H({\bf x}, {\bf y})+b-1, n\} \leq d_b({\bf x}, {\bf y})$, (see Proposition 9 in \cite{YBS16}), it follows that $B_b({\bf x}, r) \subset B_H({\bf x}, r-b+1)$. The second bound $R_b({\bf C}) \leq bR_H({\bf C})$ follows from the inequality $d_b({\bf x}, {\bf y}) \leq bd_H({\bf x}, {\bf y})$.\\

In \cite{Litsyn}, the covering radii of many known binary covering codes were determined. Then their covering radii in the pair metric are ranging from $R_H+1$ to $2R_H$. It is interesting to determine their covering radii in the pair metric exactly.\\

Let ${\bf S}_n$ be the permutation group of the order $n!$ of all $n$ coordinate permutations. For any given code ${\bf C} \subset {\bf F}_q^n$ and a permutation $s \in {\bf S}_n$, the code after permutation is denoted by $s({\bf C})$. If ${\bf C}$ is a linear $[n, k]_q$ code, then $s({\bf C})$ is also a linear $[n, k]_q$ code for any permutation $s \in {\bf S}_n$. We recall that the covering radius in the Hamming metric of a linear $[n, k]_q$ code is at most $n-k$, this is the redundancy bound, see page 217 of \cite{CHLL}.\\

{\bf Theorem 2.1.} {\em Let ${\bf C} \subset {\bf F}_q^n$ be a linear $[n, k]_q$ code, then there exists a permutation $s\in {\bf S}_n$ such that the  covering radius in the $b$-symbol metric of $s({\bf C})$ is at most $n-k+b-1$, that is, $$R_{b}(s({\bf C})) \leq\min\{ n-k+b-1, n\}.$$}\\

{\bf Proof.} Suppose that the first $k$ columns of one generator matrix are linearly independent. Then for any given vector ${\bf y}$ in ${\bf F}_q^k$ there is a codeword ${\bf c}$ in ${\bf C}$ such that the first $k$ coordinates of ${\bf c}$ equal to ${\bf y}$. Therefore in each coset ${\bf y}+{\bf C}$ we can find a vector with the first $k$ coordinates equal to zero. Then the smallest $b$-symbol weight in each coset is at most $n-k+b-1$.  If the first $k$ columns of this linear code are not linearly independent, suppose that the columns $i_1<i_2<\cdots<i_k$ are linearly independent. We take a permutation $s$ transforming $i_1, \ldots, i_k$ to $n-k+1, \ldots, n$. Then the conclusion follows immediately.\\

{\bf Corollary 2.1 (redundancy bound).} {\em Let ${\bf C}$ be a linear $[n, k]_q$ code, if there are $k$ consecutive coordinate positions $i, i+1, \ldots, i+k-1$ such that columns in one generator matrix of ${\bf C}$ at these positions are linear independent, then $R_b({\bf C}) \leq \min\{n-k+b-1, n\}$.}\\

For a linear $[n, k]_q$ code ${\bf C}$, let $wt_{coset}({\bf v}+{\bf C})$ be the minimum Hamming weight among all weights of vectors in this coset. Hence $$R_H({\bf C})=\max_{{\bf v} \in {\bf F}_q^n} \{wt_{coset}({\bf v}+{\bf C}) \},$$ see Theorem 11.1.2 in \cite{HP}. Then $R_H({\bf C})$ is minimum positive integer $s$ such each nonzero vector in ${\bf F}_q^{n-k}$ can be represented as linear combinations of at most $s$ columns in its parity check matrix. Moreover such a linear combination corresponds to a vector in one coset of the smallest Hamming weight.  Then the redundancy bound about covering radii of linear codes follows immediately. This bound $R_H({\bf C}) \leq n-k$ for a linear $[n, k]_q$ code is attained when ${\bf C}$ is a Reed-Solomon code, see \cite{CHLL}. Similarly for a linear $[n, k]_q$ code ${\bf C} \subset {\bf F}_q^n$ let $wt_{b, coset}({\bf v}+{\bf C})$ be the minimum $b$-symbol weight among all weights of vectors in this coset. Then $$R_b({\bf C})=\max_{{\bf v} \in {\bf F}_q^n} \{wt_{b, coset}({\bf v}+{\bf C}) \}.$$

{\bf Example 1.} Let ${\bf C}$ be a binary linear self-dual $[2n, n, 2]_2$ code with one generator matrix $({\bf I}_n, {\bf I}_n)$, where ${\bf I}_n$ is the $n \times n$ identity matrix. Then it is easy to verify that $R_H({\bf C})=n+1$. On the other hand the vectors in cosets can be of the form $({\bf 0}, {\bf y})$, where ${\bf 0}$ and ${\bf y}$ are vectors in ${\bf F}_2^n$, Hence the smallest possible pair weight is $n+1$, and $R_2({\bf C})=n+1$. The above redundancy bound for symbol-pair code is attained. The Example 2 shows that the condition in Corollary 2.1 is necessary.\\

{\bf Example 2.} We use different coordinate ordering of binary linear $[4t, 2t, 2]_2$ code ${\bf C}_1$ with the following generator matrix.\\

$$
\left(
\begin{array}{cccccccccccccccccccc}
1&1&0&0&0&0&0&\cdots&0&0\\
0&0&1&1&0&0&0&\cdots&0&0\\
0&0&0&0&1&1&0&\cdots&0&0\\
\cdots&\cdots&\cdots&\cdots&\cdots&\cdots&\cdots&\cdots&\cdots&\cdots\\
0&0&0&0&0&0&0&\cdots&1&1\\
\end{array}
\right)
$$

This code is a self-dual code and the above generator matrix is also a parity check matrix.  Then the smallest possible pair weight vectors in cosets  is of the form $$(0110110110\ldots0110),$$ therefore the smallest possible pair weight of vectors in cosets is $3t$. We have $R_2({\bf C}_1)=3t$. \\

The Delsarte bound in \cite{Delsarte} asserts that the covering radius of a linear code in the Hamming metric is bounded from above by the number of nonzero weights of its dual. When $t=2$ we have a self-dual binary $[4, 2, 2]_2$ code ${\bf C}_2$ with four codewords, $(0000), (1100), (0011), (1111)$. The dual code is the same. Hence the nonzero pair weights of the dual code are $3$ and $4$. It is clear that the covering radius in the pair metric is $R_2({\bf C}_2)=3$. This is a counterexample to the Delsarte bound in the pair metric.\\

{\bf Example 3.} Let ${\bf C}$ be a binary linear $[6, 3, 3]_2$ code with the following parity check matrix.\\

$$
\left(
\begin{array}{ccccccccc}
1&0&0&1&1&0\\
0&1&0&0&1&1\\
0&0&1&1&0&1\\
\end{array}
\right)
$$
Since $(000111)$ is a codeword, $d_2({\bf C})=4$. Any vector in ${\bf F}_2^3$ can be represented as the sum of at most two columns in the above parity check matrix. Hence the covering radius of the above binary linear $[6, 3, 3]_2$ code is $2$. From Proposition 2.1, the covering radius of this linear $[6, 3, 3]_2$ code in the pair-metric satisfies $3\leq R_2({\bf C}) \leq 4$. If $R_2({\bf C})=3$, then in each coset ${\bf v}+{\bf C} \subset {\bf F}_2^6$, the vector of Hamming weight $2$ has to contain two consecutive support positions. This is not the case, since we can check $(1,1,1)$ has no representation as the sum of two consecutive columns in the above matrix. Therefore $R_2({\bf C})=4$ attains the above redundancy bound for symbol pair codes and the bound in Proposition 2.1.\\

The Norse bound in \cite{HKM78} claims that for a binary linear code ${\bf C}$, if its dual distance is at least $2$, that is no two columns in any generator matrix of ${\bf C}$ are linear dependent, then the covering radius $R_H({\bf C}) \leq \lfloor\frac{n}{2}\rfloor$, or see Corollary 11.2.2  in \cite{HP}. Example 3 is a counterexample to such claim in the pair metric. Hence the Norse bound on the covering radius in the $b$-symbol metric does not hold true.\\

{\bf Example 4.} Let ${\bf C}$ be the Hamming $[8, 4, 4]_2$ code with the following generator matrix. This is a binary linear code with $d_H({\bf C})=4$ with $16$ codewords. Their weights are $4$ and $8$ and this code have two nonzero weights $4$ and $8$.\\

$$
\left(
\begin{array}{cccccccccccccccccccc}
1&1&1&1&0&0&0&0\\
0&0&0&0&1&1&1&1\\
1&1&0&0&1&1&0&0\\
0&1&1&0&0&1&1&0\\
\end{array}
\right)
$$

From the Delsarte bound it is easy to verify that the covering radius of ${\bf C}$ in the Hamming metric is $R_H({\bf C})=2$. It is clear that all pair weights of codewords are in the set $\{5, 6, 7, 8\}$. We observe the vectors of weight $2$ in each coset. It is easy to verify that there are many vectors with two support positions, one support position in $\{1, 2, 3, 4\}$ and one support position in $\{5, 6, 7, 8\}$,  in some cosets. Then the smallest pair weight $$\max_{{\bf v} \in {\bf F}_q^n} \{wt_{2, coset}({\bf v}+{\bf C}) \}=4.$$ We have $R_2({\bf C})=4$. This is an example with the covering radius in the pair metric smaller than the redundancy upper bound.\\

A symbol-pair code ${\bf C} \subset {\bf F}_q^n$ satisfies the Singleton bound $$|{\bf C}| \leq q^{n-d_2+2},$$ see \cite{CJKWY13}. A code attaining this bound is called an MDS symbol-pair code. Many linear MDS codes have been constructed, see \cite{CJKWY13,KZL15,CLL2017,DG,ML22,Dinh,Dinh20}.  Similarly a linear $[n, k]_q$ code satisfying $d_b=n-k+b$ is called an MDS $b$-symbol metric code and some linear MDS $b$-symbol metric codes have been constructed, see \cite{DG,Dinh20}.\\

{\bf Corollary 2.2.} {\em Let ${\bf C}$ be a linear MDS $b_1$-symbol metric code for $1 \leq b_1 \leq n-1$, then its covering radius in the $b$-symbol metric satisfies $R_b({\bf C}) \leq \min\{n-k+b-1, n\}$.}\\

{\bf Proof.}  Let ${\bf G}$ be one generator matrix, then the last $k$ columns have to be linearly independent. Otherwise there is a nonzero codeword such that the last $k$ coordinates are zero. Then $d_{b_1}({\bf C}) \leq n-k+b_1-1$. This code ${\bf C}$ is not an MDS $b_1$-symbol metric code. Then the conclusion follows from Corollary 2.1.\\

{\bf Theorem 2.2.} {\em The covering radius of Reed-Solomon codes in the $b$-symbol metric satisfies $R_b(RS(n, k))=\min\{n-k+b-1, n\}$.}\\

{\bf Proof.} Any $k$ consecutive coordinate positions satisfy the required property in Corollary 2.2 for Reed-Solomon codes. On the other hand the covering radius of the code $RS(n, k)$ is $n-k$. The conclusion follows from the conclusion $R_H({\bf C})+b-1\leq R_b({\bf C})$ in Proposition 2.1.\\

Since there are many MDS symbol-pair or MDS $b$-symbol linear cyclic codes constructed, see for example \cite{KZL15,Dinh20}, it is an interesting open problem to determine the covering radii of these MDS codes in the pair metric or the $b$-symbol metric, as in Theorem 2.2. In Theorem 2.4 and Corollary 2.4 some lower bounds on their covering radii in the pair metric are given.\\

We recall some basic facts about algebraic geometry codes. We refer to Chapter 10 of \cite{Lint1}, Chapter 13 of \cite{HP} and \cite{TV} for the basic notations and definitions in algebraic geometry. Let ${\bf X}$ be an absolutely irreducible  projective smooth genus $g$ curve defined over ${\bf F}_q$. Let $P_1,\ldots,P_n$ be $n$ distinct rational points of ${\bf X}$ over ${\bf F}_q$. Let ${\bf G}$ be a rational divisor over ${\bf F}_q$ of degree $\deg({\bf G})$ where $2g-2 <\deg({\bf G})<n$ and $$support({\bf G}) \bigcap {\bf P}=\emptyset.$$ Let ${\bf L}({\bf G})$ be the function space associated with the divisor ${\bf G}$. The algebraic geometry function code associated with ${\bf G}$, $P_1,\ldots,P_n$ is defined by $${\bf C}(P_1,\ldots,P_n, {\bf G}, {\bf X})=\{(f(P_1),\ldots,f(P_n)): f \in {\bf L}({\bf G})\}.$$ The dimension of this code is $$k=\deg({\bf G})-g+1$$ from the Riemann-Roch Theorem, see \cite{TV}. The minimum Hamming distance is $$d_H \geq n-\deg({\bf G}).$$ The Reed-Solomon codes are just the algebraic-geometric codes over the genus $0$ curve.  Algebraic geometry residual code with the dimension $k=n-m+g-1$ and minimum Hamming distance $d_H \geq m-2g+2$ were also defined, we refer to \cite{Lint1,HP,TV} for the detail.\\

From the Riemann-Roch Theorem, any $k-1$ columns in one generator matrix of an algebraic-geometric $[n, k]_q$ code are linear independent. Hence we have the following upper bound on the covering radius of an algebraic-geometric code.\\

{\bf Corollary 2.3.} {\em Let ${\bf C}$ be an algebraic-geometric $[n, k]_q$ code, then its covering radius in the $b$-symbol metric is at most $n-k+b-1$.}\\

{\bf Proof.} This conclusion follows from a similar argument as the proof of Theorem 2.1.\\

Some lower bounds on the covering radii in the $b$-symbol metric of linear codes are given in Theorem 2.3, 2.4 and Corollary 2.4.\\

{\bf Theorem 2.3.} {\em Let ${\bf C}$ be a linear $[n, k]_q$ code, then its covering radius in the $b$-symbol metric is at least $$(b+1) \cdot \left \lfloor \frac{n}{2(k+1)}\right \rfloor$$.}\\

{\bf Proof.} Divide $\{1, 2, \ldots, n\}$ to $t= \lfloor \frac{n}{k+1}\rfloor$ parts $A_1, \ldots, A_t$, each part have consecutive $k+1$ coordinate positions. Since any $k+1$ columns in one generator matrix of ${\bf G}$ are linearly dependent. Then we can find a vector ${\bf y}$ in ${\bf F}_q^n$  such that in the coset ${\bf y}+{\bf C}$, has at least one nonzero coordinate in $A_1, \ldots, A_t$. In every two consecutive parts, the vectors in the coset has the smallest $b$-weight, only when it is of the form $(\ldots0110\ldots)$, where the first $1$ in the part $A_i$ and the second $1$ is  in the next part $A_{i+1}$. Then the conclusion follows directly.\\

Covering radii of cyclic codes including BCH codes have been studied extensively, for example, see \cite{DJ1991,CHLL}. From the calculations of pair-distances and $b$-symbol distances of some cyclic codes and constacyclic codes in \cite{Dinh,Dinh20} we can give lower bounds on the covering radii of these codes in the pair or the $b$-symbol metric. The main point is the natural inclusion relation of these cyclic and constacyclic codes. It is obvious that similar lower bounds on the covering radius of some other cyclic or constacyclic codes can be obtained.\\

{\bf Theorem 2.4.} {\em Let $p$ be an odd prime. Let ${\bf C}_{i, i}$ be the cyclic code over ${\bf F}_{p^m}$ of the length $2p^s$ generated by $(x^2-1)^i$. Suppose that $p^s-p^{s-e}+\tau p^{s-e-1}+2 \leq i \leq p^s-p^{s-e}+(\tau+1) p^{s-e-1}+1$, where $0 \leq \tau \leq p-2$, $0\leq e \leq s-1$. Then $$R_2({\bf C}_{i, i}) \geq 2(\tau+1)p^e.$$}\\

{\bf Proof.} The cyclic codes ${\bf C}_{i,i}$ and ${\bf C}_{i-1, i-1}$ satisfy that ${\bf C}_{i, i}\subset {\bf C}_{i-1, i-1}$. Moreover these two codes are not the same. Then in the coset ${\bf v}+{\bf C}_{i,i}$ for some ${\bf v} \in {\bf C}_{i-1, i-1}\backslash {\bf C}_{i, i}$, the pair weight of each vector is at least the minimum pair-distance $d_2({\bf C}_{i-1, i-1})$ of the code ${\bf C}_{i-1,i-1}$. The conclusion follows from Theorem 9 in \cite{Dinh}.\\

Let $p$ be a prime and $s, m$ be two positive integers, $s=r_1m+r$, where $r_1$ and $r$ are two nonnegative integers satisfying $0 \leq r\leq m-1$. Let $\lambda$ be a nonzero element $\lambda \in {\bf F}_{p^m}$. Set $\gamma=\lambda^{p^{(r_1+1)m-s}}$. Then length $p^s$ $\lambda$-constacyclic code ${\bf C}_i$ is generated by $(x-\gamma)^i$, $0 \leq i \leq p^s$, see \cite{Dinh20} page 2. The following result follows from Theorem 9 in \cite{Dinh20}.\\

{\bf Corollary 2.4.} {\em Let $p$ be an odd prime. Suppose that $i$ satisfies $p^s-p^{s-e}+\tau p^{s-e-1}+\beta+1 \leq i \leq p^s-p^{s-e}+(\tau+1)p^{s-e-1}+1$ where $0 \leq e \leq s-2$, $0 \leq \tau \leq p-2$ and $b \leq \beta(\tau+1)$. Then the covering radius of ${\bf C}_i$ in the $b$-symbol metric satisfies $$R_b({\bf C}_i) \geq b(\tau+2)p^e.$$}\\

\section{Perfect $b$-symbol metric code}

When $b=2$, $B_2({\bf x}, 2)$ is exactly the $B_H({\bf x}, 1)$, we refer to \cite{CB11}, Example 2. Hence binary Hamming  $[2^n-1, 2^n-1, 3]_2$ code with the pair distance $5$ is a perfect pair metric code as showed in Theorem 19 in \cite{CB11}. It is interesting to study the perfect $b$-symbol metric codes in general.\\

{\bf Proposition 3.1.} {\em $B_b({\bf x}, r) \subset B_H({\bf x}, r-b+1)$ and $B_H({\bf x}, r) \subset B_b({\bf x}, br)$. Moreover $B_b({\bf x}, r)={\bf x}$ for $r \leq b-1$, $B_b({\bf x}, b)=B_H({\bf x}, 1)$.}\\

{\bf Proof.} This is a direct analysis about the shape of the balls.\\

The following result is a direct generalization of Theorem 19 in \cite{CB11}.\\

{\bf Proposition 3.2.} {\em The minimum $b$-symbol distance of the Hamming $[\frac{q^m-1}{q-1}, m, 3]_q$ code over ${\bf F}_q$ is at most $2b+1$. If the minimum $b$-symbol distance of this $q$-ary Hamming code is $2b+1$, it is a perfect $b$-symbol metric code.}\\

{\bf Proof.} It is clear that there is one Hamming weight $3$ codeword $${\bf c}=(110\ldots0x\ldots0)$$ in the $q$-ary Hamming code for some nonzero $x \in {\bf F}_q$. Then from the definition of $b$-symbol distance, we have $d_b({\bf c}, {\bf 0}) \leq 2b+1$. The second conclusion follows from Proposition 3.1.\\

{\bf Theorem 3.1.} {\em For any given finite field ${\bf F}_q$, there is no perfect symbol-pair code with the cardinality $q^k$ and the minimum pair-distance $7$.}\\

{\bf Proof.} We analyze the shape of the ball $$B_2({\bf 0}, 3)=B_2({\bf 0}, 2) \bigcup \{{\bf x}: d_2({\bf x}, {\bf 0})=3\}.$$ It is clear that $B_2({\bf 0}, 2)$ and the sphere $\{{\bf x}: d_2({\bf 0}, {\bf x})=3\}$ are disjoint. For any nonzero ${\bf x}$ satisfying $d_2({\bf x}, {\bf 0}) \leq 2$,  the vector ${\bf x}$ has to be the form $(0\ldots0*0\ldots0)$ with only one nonzero coordinate. When ${\bf x}$ satisfies $d_2({\bf x}, {\bf 0})=3$, it has to be the form  $(0\ldots0**0\ldots0)$ or $(*0\ldots0*)$,  with two consecutive nonzero coordinates. Therefore for each ${\bf x} \in B_2({\bf 0}, 3)$, we have $d_H({\bf 0}, {\bf x})=1$, or $d_H({\bf x}, {\bf 0})=2$ and ${\bf x}$ has two consecutive support positions. Hence the cardinality of the ball $B_2({\bf 0}, 3)$ is $|B_2({\bf 0}, 3)|=1+n(q-1)+n(q-1)^2$. \\

If there is a perfect symbol-pair code with the minimum pair-distance $7$ and the cardinality $q^k$, $|B_2({\bf 0}, 3)|q^k=q^n$. Then $\frac{q^{n-k}-1}{q-1}=n+n(q-1)$. Hence $nq=q^{n-k-1}+q^{n-k-2}+\cdots+q+1$, $nq \equiv 1$ $mod$ $q$. This is a contradiction.\\

{\bf Theorem 3.2.} {\em If ${\bf C}$ is a perfect $b$-symbol metric code, then $$d_b({\bf C}) \geq d_H({\bf C})+2b-1.$$ In particular if $b\geq \frac{n+1}{2}$, there is no perfect $b$-symbol metric code.}\\

{\bf Proof.} From Proposition 2.1, $R_b({\bf C})\geq R_H({\bf C}) +b-1\geq \lfloor\frac{d_H({\bf C})-1}{2}\rfloor+b-1$, we have $$\lfloor (d_b({\bf C})-1)/2\rfloor \geq \lfloor (d_H({\bf C})-1)/2\rfloor+b-1,$$ since for a perfect $b$-symbol metric code $R_b({\bf C})=\lfloor\frac{d_b({\bf C})-1}{2}\rfloor$. Then the conclusion follows immediately.\\

{\bf Theorem 3.3.} {\em Let ${\bf C}$ be a linear $[n, k]_q$ code. Then ${\bf C}$ is not perfect in the $(2k+2)$-symbol metric.}\\

{\bf Proof.} From the proof of Theorem 2.3, there is a vector in some coset, every consecutive $(2k+2)$ coordinate positions can not be zero. Therefore $R_{2(k+1)}({\bf C})=n$ from $R_{2(k+1)}({\bf C})=\max_{{\bf v} \in {\bf F}_q^n} \{wt_{2(k+1), coset}({\bf v}+{\bf C}) \}$. The conclusion follows immediately.\\

{\bf Theorem 3.4.} {\em Let ${\bf C} \subset {\bf C}' \subset {\bf F}_q^n$ be two linear codes satisfying $d_2({\bf C}') \geq \frac{b}{2} \cdot d_H({\bf C})$ and ${\bf C} \neq {\bf C}'$. Then ${\bf C}$ is not perfect in the $b$-symbol metric.}\\

{\bf Proof.} Let ${\bf v} \in {\bf C}'\backslash {\bf C}$, then any vector in the coset ${\bf v}+{\bf C}$ has its $b$-symbol weight at least $d_b({\bf C}')$. Therefore $R_b({\bf C}) \geq d_b({\bf C}') \geq \frac{b}{2}d_H({\bf C}) >\frac{d_b({\bf C})-1}{2}$. The conclusion follows directly.\\

The following result Corollary 3.1 follows from Theorem 3.4 for the pair metric case.\\

{\bf Corollary 3.1.} {\em Let ${\bf C}$ be an MDS symbol-pair $[n, k]_q$ linear code, then any linear subcode ${\bf C}'$ in ${\bf C}$ of dimension $k-1$ is not perfect in the pair metric.}\\

{\bf Proof.} From the Singleton bound in the Hamming metric $d_H({\bf C}') \leq n-(k-1)+1=n-k+2=d_2({\bf C})$. The conclusion follows from Theorem 3.4 immediately.\\

Similarly we can prove the following result.\\

{\bf Corollary 3.2.} {\em Let ${\bf C}$ be a linear subcode of a linear code ${\bf C}_1$ and assume that these two codes are not the same. Assume $d_2({\bf C}_1) \geq d_H({\bf C})$ or $d_H({\bf C}_1) \geq d_H({\bf C})-1$. Then ${\bf C}$ is not perfect in the pair metric.}\\

Theorem 3.1 and Corollary 3.2 indicate that many linear codes are not perfect in the pair metric. Theorem 3.4 and Corollary 3.2 can be used to exclude a lot of linear codes as perfect codes in the pair metric. For example from the computation of optimum distance profiles (ODPs) of self-dual binary codes in \cite{FK14,Freibert}, many codimension $1$ linear subcode of some self-dual binary codes are not perfect in the pair metric. Moreover a lot of  cyclic codes and algebraic-geometric codes are not perfect in the $b$-symbol metric, as proved in the following two results. We refer to \cite[Chapter 13]{HP} for the notations of algebraic geometry codes.\\

{\bf Corollary 3.3.} {\em Let ${\bf C}$ be an algebraic geometry residual code with the dimension $n-m+g-1$ and the minimum Hamming distance $m-2g+2$, defined by a degree $m$ divisor of the form $m{\bf Q}$, where $m>2g-2$ and ${\bf Q}$ is a rational point of the curve. Then ${\bf C}$ is not perfect in the pair metric.}\\

{\bf Proof.} Let ${\bf C}_1$ be the residual code defined by the divisor $(m-1){\bf Q}$. Then ${\bf C}$ is a real subcode of ${\bf C}_1$. Since $d_H({\bf C}_1) \geq m-1-2g+2 \geq d_H({\bf C}) +1$, the conclusion follows from Corollary 3.2.\\

The Hermitian curve $x^qz+z^qx=y^{q+1}$ over ${\bf F}_{q^2}$ is a genus $g=\frac{q^2-q}{2}$ curve and it has $1+q^3$ rational points. Let $m$ be a positive integer satisfying $2g-2 <m\leq q^3-1$. The Hermitian code ${\bf C}_m$ is a linear $[q^3, q^3-m+g-1, \geq m-2g+2]_{q^2}$ code, see \cite{MR20}. When $m \geq 2q^2-2q-2$, the true minimum Hamming distance is $m-2g+2$.  From Corollary 3.3 these Hermitian codes are not perfect symbol-pair codes.\\

From Theorem 8 and Theorem 9 in \cite{Dinh} most cyclic codes of the length $2p^s$ over ${\bf F}_{p^m}$ of the form ${\bf C}_{i, i}$ as in Section 2 can not be perfect. We give the following result.\\

{\bf Corollary 3.4.} {\em Let $p$ be an odd prime. Let ${\bf C}_{i, i}$ be the cyclic code over ${\bf F}_{p^m}$ of the length $2p^s$ generated by $(x^2-1)^i$. Suppose that $p^s-p^{s-e}+\tau p^{s-e-1}+2 \leq i \leq p^s-p^{s-e}+(\tau+1) p^{s-e-1}+1$, where $0 \leq \tau \leq p-2$, $0\leq e \leq s-1$. Then ${\bf C}_{i, i}$ can not be a perfect code in the pair metric.}\\

{\bf Proof.} From Theorem 8 and Theorem 9 in \cite{Dinh}, the pair-distance of the code ${\bf C}_{i,i}$ is $d_2({\bf C}_{i, i})=2d_H({\bf C}_{i-1, i-1})$. The conclusion follows from Corollary 3.2.\\

Similarly many constacyclic codes studied in \cite{Dinh20} are not perfect in the $b$-symbol metric from Corollary 3.2, and the exact determination of their minimum Hamming distances and minimum pair-distances in \cite{Dinh20}. It is an interesting problem to determine all perfect symbol-pair codes.\\

\section{The generalized Singleton bound}

When $b=n-1$, $B_{b-1}({\bf x}, n-1)=B_H({\bf x}, 1)$, hence any code is $(n-1, (1+n(q-1)))$ list-decodable $(n-1)$-symbol metric code. It is not interesting to discuss list-decodable $b$-symbol metric codes when $b$ is big.\\

From the definitions above, for an $(d_{list}, L)$ list-decodable code ${\bf C} \subset {\bf F}_q^n$  in the $b$-symbol metric, and a given covering code ${\bf C}_1 \subset {\bf F}_q^n$ in the $b$-symbol metric with the covering radius at most $d_{list}$, we have the following upper bound on the size of the list-decodable code or the lower bound on the list size, $$|{\bf C}| \leq L|{\bf C}_1|.$$ Actually balls centered at codewords of ${\bf C}_1$ of the radius $d_{list}$ cover the whole space ${\bf F}_q^n$. Then the number of all codewords in ${\bf C}$ is not bigger than the number of codewords in these $|{\bf C}_1|$ balls.  From this simple observation and some covering codes in the $b$-symbol metric, the generalized Singleton bound on list-decodable $b$-symbol metric codes can be obtained. Notice that in the above observation, we can take any covering code ${\bf C}_1$ satisfying $R_b({\bf C}_1) \leq d_{list}$. Hence the key point in the following bounds in Proposition 4.1 and Theorem 4.1 is to find the small covering code ${\bf C}_1$ with the covering radius in the $b$-symbol metric $R_b({\bf C}_1) \leq d_{list}$.\\

{\bf Proposition 4.1.} {\em An $(d, L)$ list-decodable $b$-symbol metric code ${\bf C} \subset {\bf F}_q^n$ has its cardinality $|{\bf C}| \leq L \cdot q^{n-d+b-1}$.}\\

{\bf Proof.} We take a linear $[n, n-d+b-1]_q$ code ${\bf C}_1$ with linearly independent first $k$ columns in one of its generator matrix. This is the covering code in the $b$-symbol metric used in our above observation. It is clear that the covering radius of ${\bf C}_1$ in the $b$-symbol metric is at most $n-(n-d+b-1)+b-1=d$. The conclusion follows from the above simple observation.\\

When $L=1$ and $d=\lfloor(d_b({\bf C})-1)/2\rfloor$, the bound in Proposition 4.1 is weaker than the Singleton bound $|{\bf C}| \leq q^{n-d_b({\bf C})+b}$.  However Proposition 4.1 holds for any list size.\\

The covering code ${\bf C}_1$ used in the proof of Proposition 4.1 is not small enough, hence the bound in Proposition 4.1 can be improved significantly if a smaller covering code ${\bf C}_1$ in the $b$-symbol metric is used. We have the following result.\\

{\bf Theorem 4.1.} {\em Let $q$ be a prime power, $m$ be a positive integer, $t$ be an even positive integer, and set $n=\frac{t(q^m-1)}{q-1}$. Let ${\bf C} \subset {\bf F}_q^n$ be a $(bt, L)$ list-decodable $b$-symbol code, where $b \leq \frac{q^m-1}{q-1}$. Then $$|{\bf C}| \leq L \cdot q^{n-\frac{md}{b}}.$$}\\

{\bf Proof.}  Let ${\bf H}$ be the following $tm \times n$ matrix where $n=\frac{t(q^m-1)}{q-1}$.\\

$$
\left(
\begin{array}{cccccccccccccccc}
{\bf H}_1&{\bf 0}&{\bf 0}&\cdots&{\bf 0}\\
{\bf 0}&{\bf H}_2&{\bf 0}&\cdots&{\bf 0}\\
{\bf 0}&{\bf 0}&{\bf H}_3&\cdots&{\bf 0}\\
\cdots&\cdots&\cdots&\cdots&\cdots\\
{\bf 0}&{\bf 0}&{\bf 0}&\cdots&{\bf H}_t\\
\end{array}
\right)
$$

Here ${\bf H}_i$ are the parity check $m \times \frac{q^m-1}{q-1}$ matrix for the $[\frac{q^m-1}{q-1}, \frac{q^m-1}{q-1}-m]_q$ Hamming code. Let ${\bf C}_1$ be the linear  $[n=\frac{t(q^m-1)}{q-1}, k=\frac{t(q^m-1)}{q-1}-tm]_q$ code with the above parity check matrix ${\bf H}$. This is the covering code in the $b$-symbol metric used in our above observation. From the equality $$R_b({\bf C}_1)=\max_{{\bf v} \in {\bf F}_q^n} \{wt_{b, coset}({\bf v}+{\bf C}_1) \},$$ then $R_b({\bf C}_1) \leq bt=\frac{b(n-k)}{m}$ if $b \leq q^m-1$.\\

The balls in the $b$-symbol metric centered at codewords of ${\bf C}_1$ with the radius $bt$ cover the whole space, since $R_b({\bf C}_1) \leq bt$. Each such ball contains at most $L$ codewords of ${\bf C}$. The conclusion follows directly.\\

Theorem 4.1 improves Proposition 4.1 significantly since we take a better auxiliary covering code ${\bf C}_1$. Generally if a covering code with the smaller size can be found and used in the above simple observation, a better generalized Singleton bound can be  obtained.\\

When $\frac{m}{b}$ is larger than $2+\epsilon$ and $b$ is small (for example $b=2$), where $\epsilon$ is any small positive real number, this bound is better than the Singleton bound $|{\bf C}| \leq q^{n-d_b({\bf C})+b}$ for the list-decodable codes with list size $L=1$ (and then $d=\left \lfloor \frac{d_b({\bf C})-1}{2}\right \rfloor$), since we have $\frac{m}{b} d=\frac{m}{b}\left \lfloor \frac{d_b({\bf C})-1}{2}\right \rfloor > d_b({\bf C})$. This is the generalized Singleton bound for list-decodable $b$-symbol codes, which can be argued simply from a covering $b$-symbol metric code.\\

{\bf Corollary 4.1.} {\em Let $q$ be a prime power and $n$ be a positive integer satisfying $n=\frac{t(q^5-1)}{q-1}+v$, where $t$ is a positive integer and $0\leq v \leq \frac{q^5-1}{q-1}-1$. Let ${\bf C} \subset {\bf F}_q^n$ be a length $n$ symbol-pair code with the minimum pair distance $D \geq 4t+1$. Then $$|{\bf C}| \leq q^{n-5t}.$$.}\\

{\bf Proof.} We take the auxiliary covering code ${\bf C}_2={\bf C}_1 \times {\bf F}_q^v$ in the above observation. Then $R_2({\bf C}_2) \leq 2t$. Then balls in the pair-metric centered at codewords of ${\bf C}_2$ with the radius $2t$ cover the whole space. Moreover each such ball centered at one codeword of ${\bf C}_2$ with the radius $2t \leq \frac{D-1}{2}$ contains at most one codeword of ${\bf C}$. Then $|{\bf C}| \leq |{\bf C}_2|=q^{n-5t}$.  The conclusion is proved.\\

The above results assert that for a given minimum pair-distance when the code length is long, the above generalized Singleton bound is much stronger than the Singleton bound $|{\bf C}| \leq q^{n-d_2({\bf C})+2}$ in \cite{CJKWY13} on the sizes of the symbol-pair codes.\\

Since the construction of MDS symbol-pair codes in two papers \cite{CJKWY13,KZL15}, there have been many constructions of the MDS symbol-pair codes in \cite{CLL2017,DG,Dinh,Dinh20,ML22} from cyclic or constacyclic codes. However no upper bound on the lengths of general MDS symbol-pair codes has been given. It was proved in \cite{DG} that lengths of linear MDS symbol-pair codes over ${\bf F}_q$ of the minimum pair distance $5$ or $6$ can not be larger than $q^2+q+1$ or $q^2$. From Corollary 4.1 we give the following upper bound on the lengths of (even no linear) MDS symbol-pair codes.\\

{\bf Corollary 4.2.} {\em Let ${\bf C} \subset {\bf F}_q$ be a symbol-pair code with the length $n$ and the minimum pair distance $D$. If $n$ and $D$ satisfy $D>17$ and $n>\frac{(D-1)(q^5-1)}{4(q-1)}$. Then the symbol-pair code ${\bf C}$ is not MDS. In particular there is no MDS symbol-pair binary code with the minimum pair distance $D>17$ and the length $n\geq \frac{31(D-1)}{4}$.}\\

{\bf Proof.} Set $t=\left \lfloor \frac{D-1}{4} \right \rfloor$. If $n \geq \frac{(D-1)(q^5-1)}{4(q-1)}$ and $D>17$, then $|{\bf C}| \leq q^{n-5t} <q^{n-D+2}$ from Corollary 4.1. The conclusion is proved.\\

If smaller covering codes in the $b$-symbol metric could be found, then Theorem 4.1 and Corollary 4.2 could be improved. To the best of our knowledge, Corollary 4.2 is the first general upper bound on the lengths of MDS symbol-pair codes.\\

\section{Conclusion}
In this paper covering codes and covering radii of some famous linear codes in the $b$-symbol metric are considered. First of all some results about covering codes in the Hamming metric, such as the Delsarte bound and the Norse bound, do not hold in the pair metric. Some highly nontrivial upper and lower bounds on covering radii of some cyclic, constacyclic and algebraic geometry codes in the $b$-symbol metric are given. The covering radius of the Reed-Solomon code as a $b$-symbol metric code is determined. As an application, the generalized Singleton bound on list-decodable $b$-symbol metric codes is proved and an upper bound on the lengths of general MDS symbol-pair codes is given. We give a simple sufficient condition for non-perfect code in the $b$-symbol metric and prove that many well-known codes are not perfect in the $b$-symbol metric. It is an interesting open problem to classify all perfect codes in $b$-symbol metric for small $b=2$ or $3$. In particular is there any perfect symbol-pair code of the minimum pair-distance bigger than or equal $9$?\\


\begin{thebibliography}{10}

\bibitem{BR17} E. Byrne and A, Ravagnani, Covering radius of matrix codes endowed with rank metric, SIAM J. Discrete Math., vol. 31, no. 2, pp. 927-944, 2018.


\bibitem{CB11} Y. Cassuto and M. Blaum, Codes for symbol-pair read channels, IEEE Trans. Inf. Theory, vol. 57, no. 12, pp. 8011-8020, 2011.

\bibitem{CL2011} Y. Cassuto and S. Litsyn, Symbol-pair codes: Algebraic constructions and asymptotic bounds, Proc. IEEE Int. Sym. Inf. Theory Process., St. Peterburg, pp. 2348-2352, Jun. 2011.

\bibitem{CJKWY13} Y. M. Chee, L. Ji, H. M. Kiah, C. Wang and J. Yin, Minimum distance separable codes for symbol-pair read channels, IEEE Trans. Inf. Theory, vol. 59,  no. 11, pp. 7259-7267, 2013.

\bibitem{CLL2017} B. Chen, L. Lin and H. Liu, Constacyclic symbol-pair codes: Lower bounds and optimal constructions, IEEE Trans. Inf. Theory, vol. 63,  no. 12, pp. 7661-7666, 2017.
\bibitem{KChen} K. Chen, On the non-existence of perfect codes with rank distance, Math. Nachr., vol. 182, pp. 89-98, 1996.

\bibitem{CHLL} G.  D. Cohen, I. Honkala, S. Litsyn and A. Lobstein, Covering codes, North-Hollan Math, Libarary, Elsecier, 1997.

\bibitem{Delsarte} P. Delsarte, Four fundamental parameters of a code and their combinatorial significance, Inform. and Control, vol. 23, pp.407-438, 1973.

\bibitem{Dinh} H. Q. Dinh, B. T. Nguyen and S. Sriboonchitta, MDS symbol-pair cyclic codes of length $2p^s$ over ${\bf F}_{p^m}$,  IEEE Trans. Inf. Theory, vol. 66,  no. 1, pp. 240-262, 2020.



\bibitem{Dinh20} H. Q. Dinh, X. Wang, H. Liu and S. Sriboonchitta, On the $b$-distances of repeated-root constacyclic codes of prime power lengths, Disc. Math., vol. 343, no. 4, 11780, 2020.

\bibitem{DG} B. Ding, G. Ge, J. Zhang, T. Zhang and Y. Zhang, New constructions of MDS symbol-pair codes, Des., Codes and Cryptogra., vol. 86, no. 4, pp. 841-859, 2018.

\bibitem{DJ1991} R. Dougherty and H. Janwa, Covering radius computations for binary cyclic codes, Math. Comp., vol. 57, no. 195, pp. 415-434, 1991.

\bibitem{EFS} D. Elimelech,  M. Firer and M. Schwartz, The generalized covering radii of linear codes, IEEE Trans. Inf. Theory, vol. 67,  no. 12, pp. 8070-8085, 2021.

\bibitem{EGY20} O. Elishco, R. Gabrys and E. Yaakobi, Bounds and constructions of codes over symbol-pair read channels, IEEE Trans. Inf. Theory, vol. 66,  no. 3, pp. 1385-1395, 2020.

\bibitem{FK14} F. J. Freibert and J-L. Kim, Optimal subcodes and optimum distance profiles of self-dual codes, Finite Fields Appl., vol. 25, pp. 146-164, 2014.

\bibitem{Freibert} F. J. Freibert, Self-dual codes, subcode structures and applications, University of Louisville Dissertation, 2012.

\bibitem{GY3} M. Gadouleau and Z. Yan, Packing and covering properties of rank-metric codes, IEEE
Trans. Inf.  Theory, vol. 54, no. 9, pp. 3873-3883, 2008.

\bibitem{HKM78} T. Helleseth, T. Klove and J. Mykkeltveit, On the covering radius of binary codes, IEEE Trans. Inf. Theory, vol. 24,  pp. 627-628, 1978.


\bibitem{HP} W. C. Huffman and V. Pless, Fundamentals of error-correcting codes, Cambridge University Press, Cambridge, U. K., 2003.


\bibitem{Lint} J. H. van Lint, Nonexistence theorems for perfect error-correcting codes, Computers in Algebra and Theory, vol. IV (SIAM-AMS Proceedings), 1971.


\bibitem{Lint1} J. H. van Lint, Introduction to conding theory, Springer, Berlin, Heidelberg, Paris, Tokyo, Third Revised Expanded Version, 1999.

\bibitem{LP} H. Liu and X. Pan, Generalized pair weights of linear codes and linear isopmorphisms preserving pair weights, IEEE Trans. Inf. Theory, vol. 68,  no. 1, pp. 105-117, 2022.

\bibitem{LXY1} S. Liu, C. Xing and C. Yuan, List decoding of symbol-pair codes, IEEE Trans. Inf. Theory, vol. 65,  no. 8, pp. 4815-4821, 2019.

\bibitem{Litsyn} S. Litsyn, Tables of the best known binary covering codes, http://www.eng.tau.il/~litsyn/.

\bibitem{KZL15} X. Kai, S. Zhu and P. Li, A construction of new MDS symbol-pair codes, IEEE Trans. Inf. Theory, vol. 61,  no. 11, pp. 5828-5834, 2015.

\bibitem{ML22} J. Ma and J. Luo, MDS symbol-pair codes from repeated-root cyclic codes, Des. Codes and Cryptogr., vol. 90, pp. 121-137, 2022.

\bibitem{MScode} F. J.  MacWilliams and N. J. A. Sloane, The Theory of error-correcting codes, 3rd Edition, North-Holland Mathematical Library, vol. 16. North-Holland, Amsterdam, 1977.

\bibitem{MR20} C. Marcolla and M. Roggero, Hermitian codes and complete intersections, Finite Fields Appl., vol. 62, 101621, 2020.

\bibitem{ShangguanTamo} C. Shangguan and I. Tamo, Combinatorial list-decoding of Reed-Solomon codes beyond the Johnson radius, Proc. 52th ACM Symp. on Theory of Computing (STOC), pp. 538-551, 2020.

\bibitem{TV}  M. Tsfasman and S. G. Vl\'{a}dut, Algebraic-geometric codes, Vol.58, Springer Science and Business Media, Netherland, 2013.


\bibitem{SOS21} M. Shi, F. \"{O}zbudak and P. Sol\'{e}, Geometric approach to $b$-symbol Hamming weights of cyclic codes, IEEE Trans. Inf. Theory, vol. 67,  no. 6, pp. 3735-3751, 2021.

\bibitem{Tie1973} A. Tiet\"{a}v\'{a}inen, On the nonexistence of perfect codes over finite fields, SIAM J. Appl. Math., vol. 24, pp. 88-96, 1973.

\bibitem{YBS16} E. Yaakobi, J. Bruck and P. H. Siegel, Construction and decoding of cyclic codes over $b$-symbol read channels, IEEE Trans. Inf. Theory, vol. 62,  no. 4, pp. 1541-1551, 2016.




\end{thebibliography}
\end{document}